\documentclass[pdftex]{JINST}
\pdfoutput=1

\usepackage[english]{babel}
 
\usepackage{graphicx}
\usepackage{lineno}
\usepackage{wrapfig}
\usepackage{subfigure}
\usepackage{rotating}
\usepackage{amssymb}
\usepackage{mathptmx}
\usepackage{multirow}

\usepackage[T1]{fontenc}
\usepackage{babel}
\usepackage[font=small,labelfont=bf]{caption}
\usepackage{footmisc}

\usepackage[numbers]{natbib}

\title{LHCb RICH Upgrade: an overview of the photon detector and electronics system}
\author{L. Cassina$^{ab}$, on behalf of the LHCb RICH Upgrade collaboration\\
\llap{$^a$} INFN, Sezione di Milano Bicocca,\\
Piazza della Scienza 3, 20126, Milano, Italy\\
\llap{$^b$} Dipartimento di Fisica G. Occhialini, Universit\`a degli Studi di Milano Bicocca,\\
Piazza della Scienza 3, 20126, Milano, Italy\\
E-mail: \email{lorenzo.cassina@mib.infn.it}}

\abstract{The LHCb experiment is one of the four large detectors operating at the LHC at CERN and it is mainly devoted to CP violation measurements and to the search for new physics in rare decays of beauty and charm hadrons. 
The data from the two Ring Image Cherenkov (RICH-1 and RICH-2) detectors are essential to identify particles in a wide momentum range.
From 2019 onwards 14 TeV collisions with luminosities reaching up to $2\cdot10^{33}$ cm$^{-2}$s$^{-1}$ with 25 ns bunch spacing are planned, with the goal of collecting 5 fb$^{-1}$ of data per year. 
In order to avoid degradation of the PID performance at such high rate (40 MHz), the RICH detector has to be upgraded.
New photodetectors (Multi-anode photomultiplier tubes, MaPMTs) have been chosen and will be read out using a 8-channels chip, named CLARO, designed to sustain a photon counting rate up to 40 MHz, while minimizing the power consumption and the cross-talk.
 A 128-bit digital register allows selection of thresholds and attenuation values and provides features useful for testing and debugging. 
Photosensors and electronics are arranged in basic units, the first prototypes of which have been tested in charged particle beams in autumn 2014. 
An overview of the CLARO features and of the readout electronics is presented.}

\begin{document}

\section{Introduction}
\begin{figure}[h!]
	\centering
	\includegraphics[width=1\textwidth]{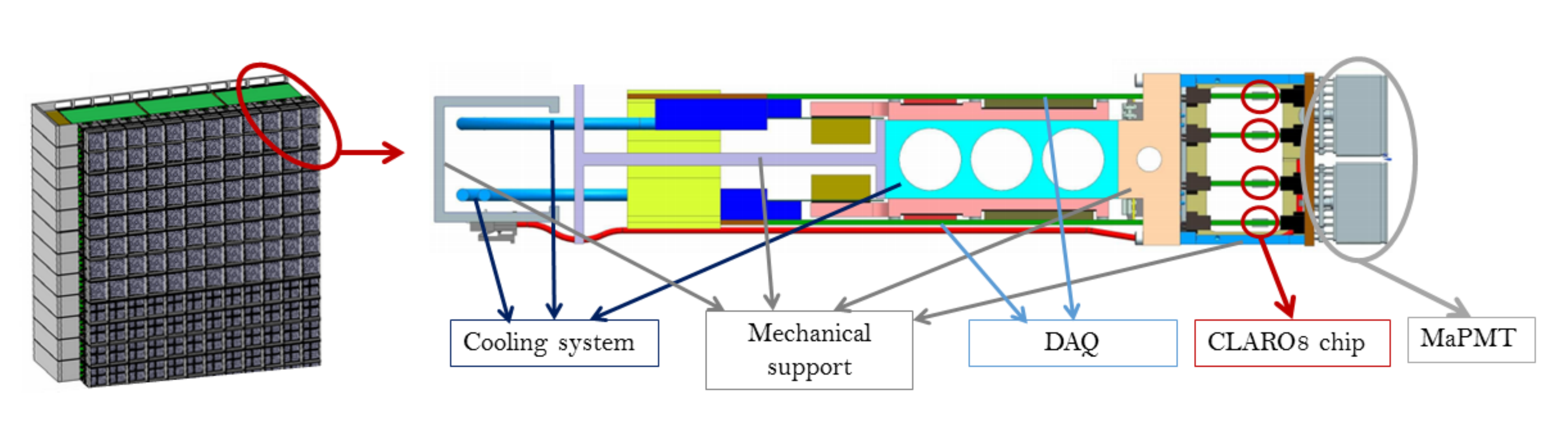}
	\caption{Schematic view of a RICH photosensitive plane, composed of modules called Elementary Cell (EC), whose structure is also shown on the right.}
	\label{fig:EC}
\end{figure}

Among the detectors operating at the Large Hadron Collider (LHC) at CERN, the LHCb experiment \cite{bib2} is designed for the research of CP violation and physics beyond the standard model by studying rare decays of  beauty and charm hadrons.  
So far, the LHCb detector has operated at a luminosity of $ 4 \times 10^{32}$~$\mathrm{cm^{-2}s^{-1}}$, although the LHC would be able to supply higher values. 
In order to make the detector capable to run at luminosities up to $ 2 \times 10^{33}$~$\mathrm{cm^{-2}s^{-1}}$ with 14 TeV proton-proton collision and 25 ns bunch spacing (providing $5$~$\mathrm{fb^{-1}}$ per year), an upgrade of the whole LHCb detector is needed \cite{LetterOfIntentUpgrade}. In particular, the Ring Image Cherenkov (RICH) system, which provides particle identifications (PID) of charged hadrons over a large momentum range, is planned to be upgraded by 2019 to allow operation at the higher luminosity and provide a 40 MHz readout rate. Although the overall structure of both RICH detectors will remain unchanged, significant modifications are required. As far as the RICH photodetectors are concerned, the Hybrid Photon Detector (HPD) used so far \cite{bib1} will be replaced by Multi-anode PhotoMultiplier Tubes (MaPMTs) coupled with external wide-bandwidth readout electronics \cite{bib3} . The baseline device that has been chosen for the RICH-1 detector is the Hamamatsu R11265 MaPMT, a 64 channel, 1 inch $\times$ 1 inch square tube capable of detect single photons \cite{NostroArticolo}. The photosensitive plane is composed of four such devices, grouped into modules called Elementary Cell (EC). As can be observed in fig.\ref{fig:EC}, the four MaPMTs are plugged in a baseboard that provides the bias and couples the anodes to four front-end boards, each hosting eight CLARO ASICs. The CLARO is a 8 channel chip realized in \hbox{0.35 $\mu$m} AMS CMOS technology \cite{CLARO} which provides a digital signal whenever the charge collected at the anode of the tube exceeds a adjustable threshold, allowing a fast photon counting. The main features of the CLARO chip will be described in section \ref{sec:CLARO}. The digital signals at the CLARO output pass through the backboard and are collected by a DAQ board. This board allows not only to configure the CLARO working parameters, but also to calibrate both the MaPMTs and the CLAROs. The two main procedures developed for this purpose will be described in section \ref{sec:Scans}.
In October and November 2014, the very first version of the system has been assembled and studied, using a charged particle beam at CERN. The Cherenkov photons were produced in a solid radiator and focussed on two EC prototypes. The system proved the capability to detect single photons and the Cherenkov rings were acquired with the expected spatial resolution. Further details and the results of the test-beam will be shown in section \ref{sec:TestBeam}. 

\section{The CLARO8 ASIC}\label{sec:CLARO}

The CLARO chip is a 8 channel custom designed ASIC realized in 0.35 $\mu$m CMOS technology from Austria Micro Systems (AMS)\footnote{AMS website: http://asic.ams.com}. Such a relatively aged and inexpensive technology can still meet the LHCb requirements in terms of wide-bandwidth and low power, while ensuring a very high yield and a good tolerance to radiation.


The CLARO chip is essentially composed of an input Charge Sensitive Amplifier (CSA) and a Discriminator. 
When a photon hits the R11265 MaPMT surface, a photoelectron is emitted from the photocathode starting the charge multiplication over 12 dynodes. The collection time of the photodetector is very small, of the order of 1 ns. Thus, the typical signal at the anode consists of a $\sim1$ Me$^-$ current pulse, injected at the input node of the CLARO. 
The CSA integrates the current pulse and provides an exponentially shaped voltage signal, whose amplitude is proportional to the input charge collected. The rise time constant of the CSA is of the order of $\sim1$ ns and is proportional to the input capacitance, while the fall time constant amounts to few ns, large enough for an effective integration of the fast pulses but short enough to sustain high rates without pile-up. The CSA is DC-coupled to a discriminator stage which provides a digital pulse if the signal at its input crosses a adjustable threshold level. The FWHM of the digital output signals is lower than 25 ns so that photon counting rates up to 40 MHz can be sustained avoiding pile-up, as required for the RICH upgrade. 
Despite its wide bandwidth, the power consumption in idle mode is low and amounts to about 1 mW per channel and it stays below 2 mW per channel even at a photon counting rate of 10 MHz. This very demanding feature is necessary to minimize the heat injection in the most illuminated areas of the RICH detector, avoiding the need for front-end cooling in such a closely packed system.
Since the response of the MaPMT pixels typically differs in gain by a factor of three \cite{NostroArticolo} \cite{H12700}, the CLARO provides attenuation and trigger threshold values that have to be adjusted to compensate the photosensor gain non-uniformity. In particular, each channel is equipped with a 12-bit register, so that the CSA attenuation (2 bits, 4 values available) and the discriminator threshold level (6 bits, 64 threshold steps of $\sim 30$ ke$^-$ each) can be correctly adjusted. As further discussed in section \ref{sec:Scans}, through the digital block the procedures developed to characterize and calibrate each channel of the system can be performed.

\section{CLARO and MaPMT calibration in the Elementary Cell}\label{sec:Scans}
Given the large number of channels ($\sim 3 \cdot 10^5$, considering both RICH detectors), automatic procedures are necessary for system calibration. Moreover, in order to periodically calibrate the system during the data taking (for instance, to compensate the drift occurring due to the loss of gain caused by the ageing of the photosensors), it is mandatory to include this functionality in the EC.

\begin{figure}[h!]
	\centering
		\subfigure[][]{\includegraphics[width=.4825\textwidth]{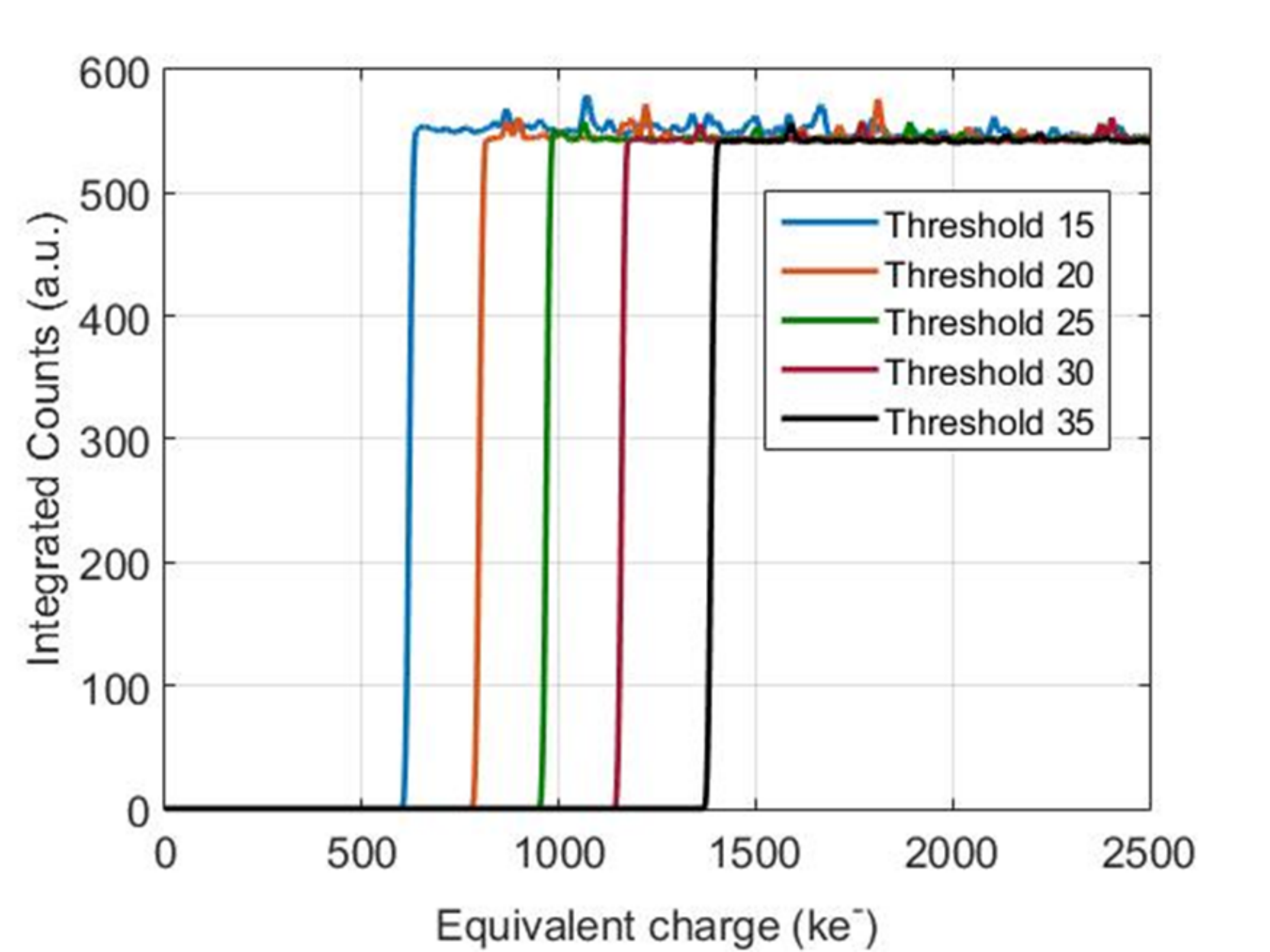}}
		\hspace{1mm}%
		\subfigure[][]{\includegraphics[width=.4825\textwidth]{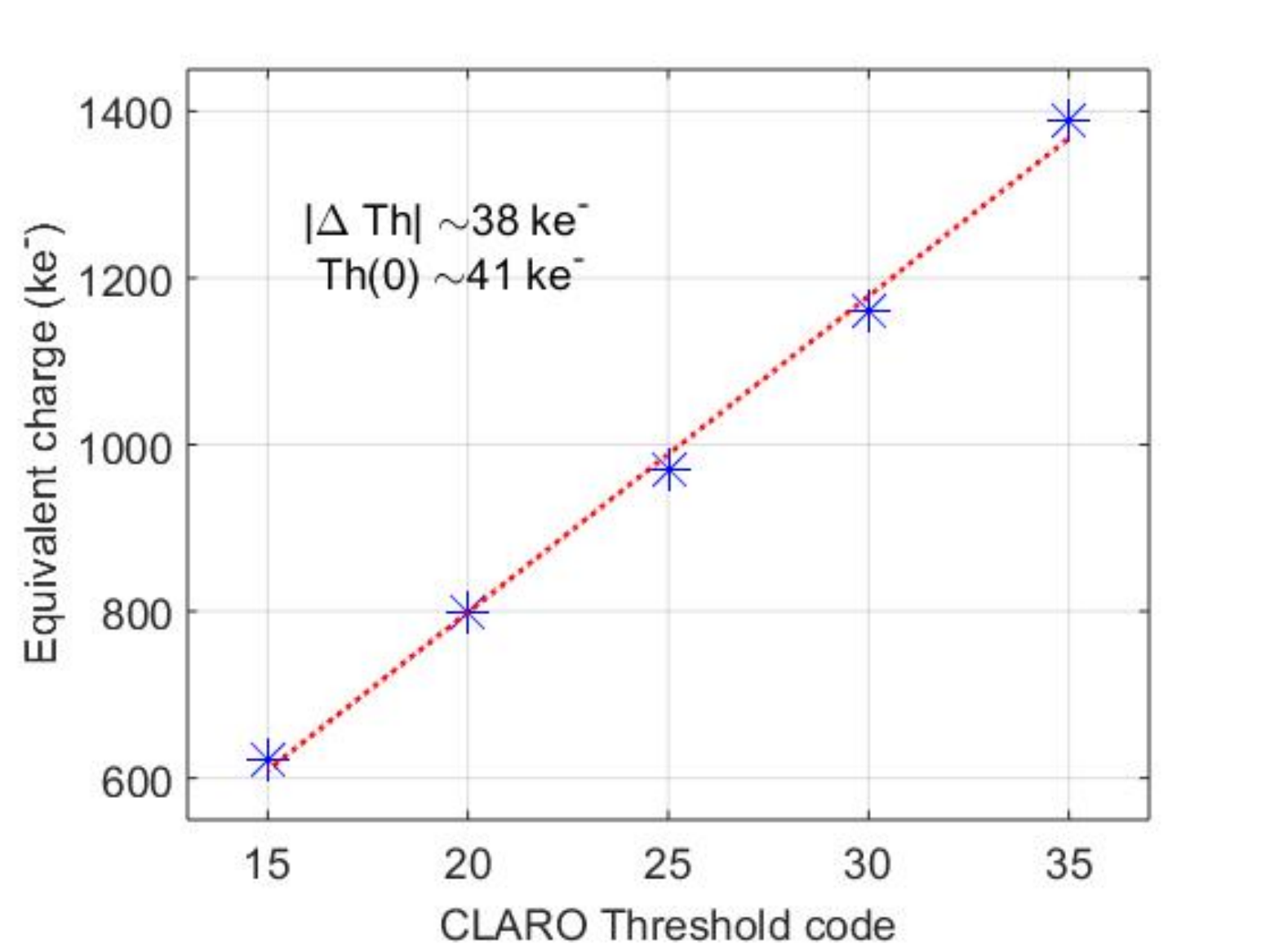}}
		\caption{Calibration of the CLARO: performing a DAC scan for various discriminator threshold levels, S-curves can be acquired. From these measurements, a linear calibration trend can be obtained measuring the offset and the threshold step amplitude of the CLARO channel.}\label{fig:DAC}
\end{figure}

The calibration of the CLARO discriminator thresholds converts each value into an equivalent amount of charge at the input. In the present version of the EC, this process exploits a commercial 10-bit Digital-to-Analog Converter (DAC) located in the EC baseboard and controlled remotely by means of the DAQ board. The DAC loads a 640 fF capacitor integrated in the CLARO ASIC with a known voltage signal. The discharge of the capacitor can be triggered using either an external signal or a CLARO self-generated pulse. As the capacitance discharges, a known charge is injected in the system. By gradually increasing the amplitude of the DAC signal, the calibrating charge ranges from a few ke$^-$ to several Me$^-$, spanning the region of interest. With a given CLARO threshold, the number of recorded signals is zero as far as the injected charge is lower than the discriminator threshold. When this value is exceeded, all the pulses injected at the input are detected. As a result, the number of recorded signals as a function of the DAC voltage value describes a S-curve. Figure \ref{fig:DAC}a shows various S-curves acquired at different discriminator threshold values. The derivative of a S-curve has a gaussian shape whose centre corresponds to the threshold equivalent input charge, while its width gives the equivalent noise charge. Repeating the process at different thresholds, the expected proportionality between the equivalent input charge and the CLARO threshold code can be obtained, as shown in fig.\ref{fig:DAC}b. The slope of the fitted line gives the mean threshold step value (design value: 30 ke$^-$) and also the offset at zero-threshold, which can vary due to the mismatches of the integrated components. If necessary, a current source can be enabled to move the zero threshold value by half the range allowing full compensation of the offset. Thus, the DAC scan gives the possibility to characterize the electronic read-out chain.

\begin{figure}[h!]
	\centering
		\subfigure[][]{\includegraphics[width=.4825\textwidth]{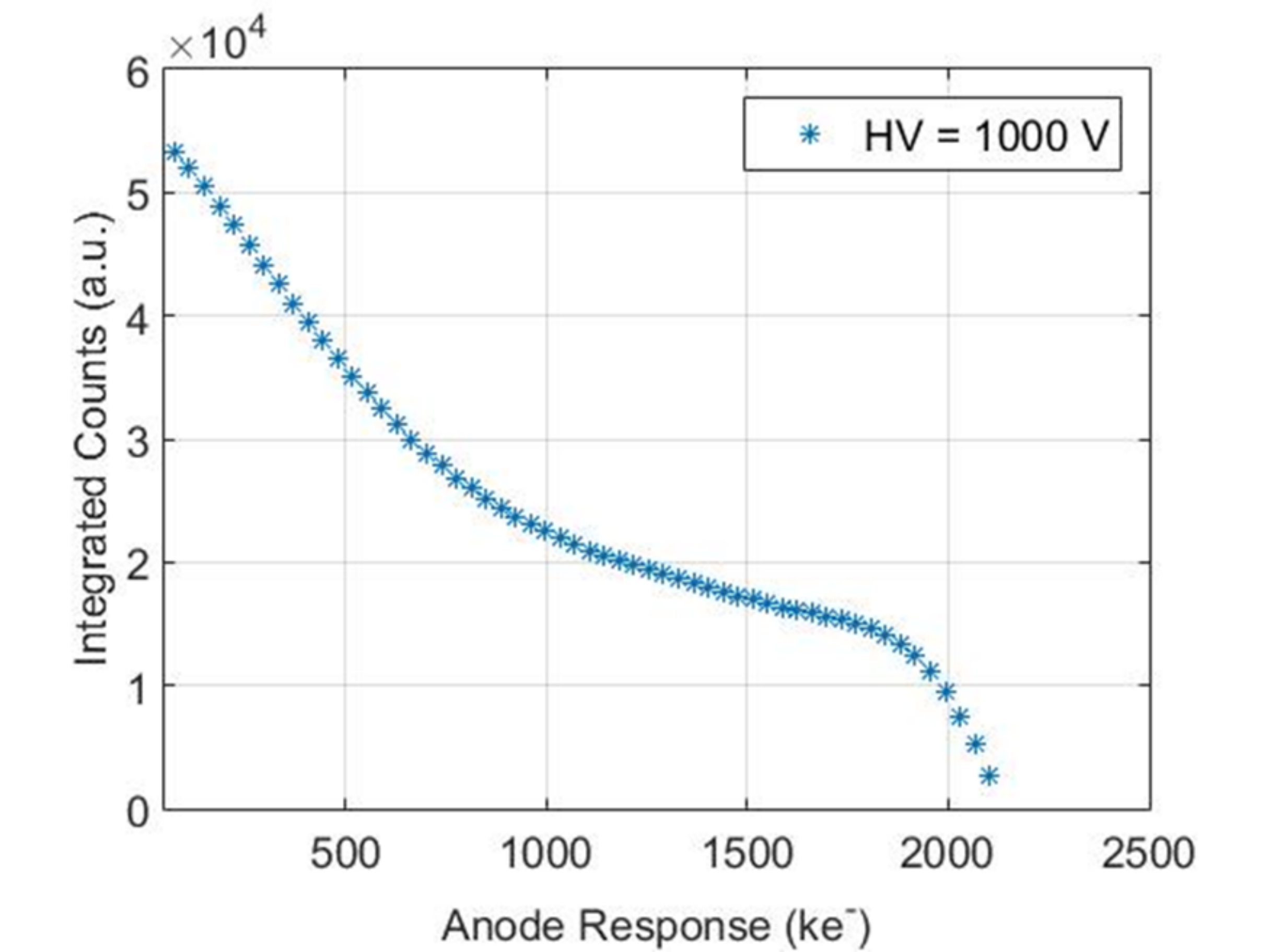}}
		\hspace{1mm}%
		\subfigure[][]{\includegraphics[width=.4825\textwidth]{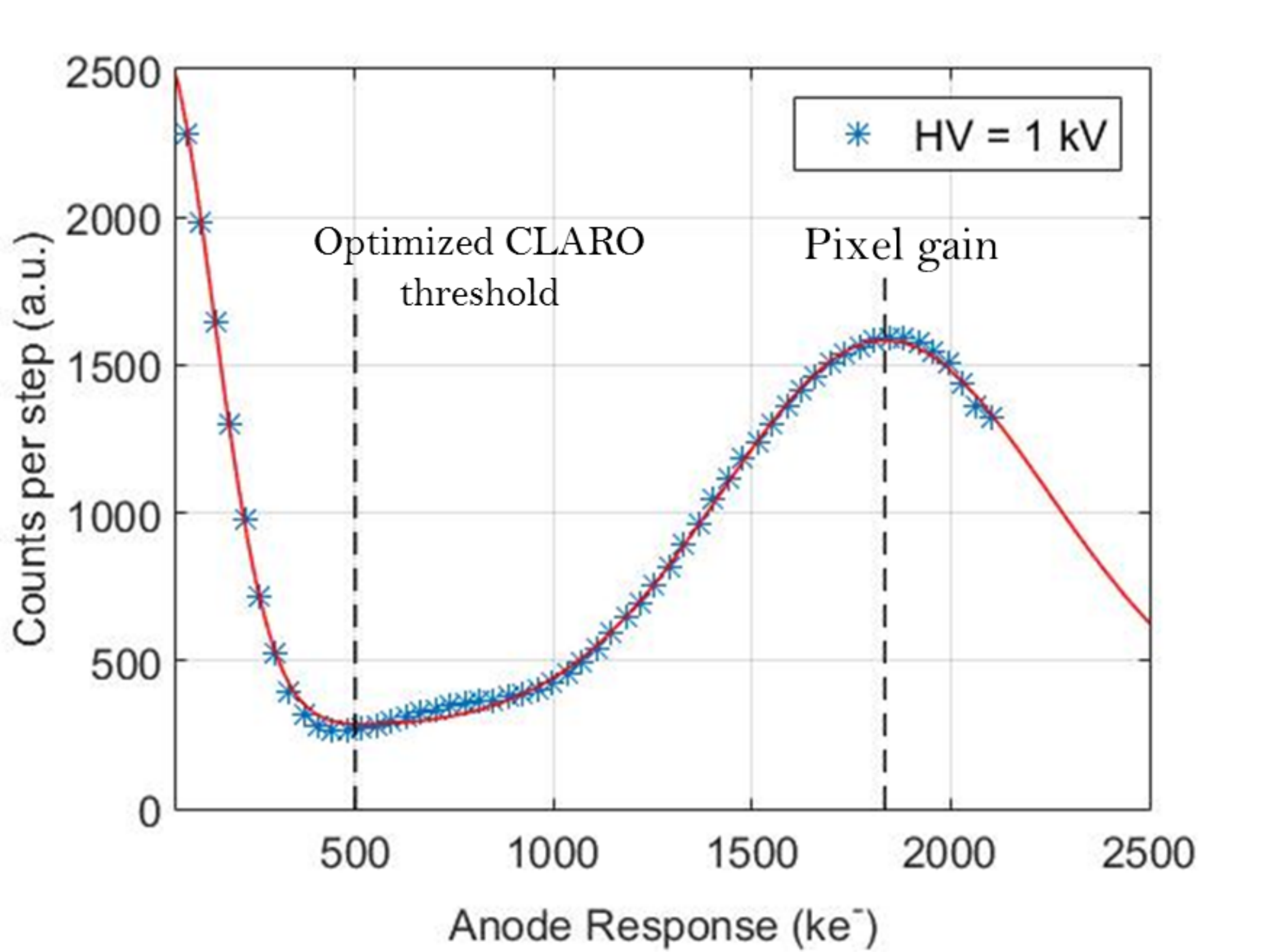}}
		\caption{Calibration of the MaPMT: by taking the derivative of the threshold scan curve, single photon spectra can be acquired. This allows to estimate the gain of each pixel and to optimize the threshold pixel-by-pixel in order to reject spurious counts. Note that the x-axis is expressed in equivalent charge, after performing a DAC scan.}\label{fig:SinglePhoton}
\end{figure}

The status of the photosensor can be studied by acquiring the single photon spectrum of the pixel under test, i.e. the distribution of the amplitude of the signals that reach the CLARO input. The single photon peak position can be used to estimate the gain of the pixel and its drift as a function of temperature and time \cite{NostroArticolo} \cite{H12700}. In addition, since the CLARO provides only digital information, it is important to choose a threshold level which maximizes the signal-to-noise ratio. One way to achieve this is to set the threshold level at the minimum of the valley between the low-amplitude pedestal (mainly populated by noise signals due to spurious electron emission from the dynodes) and the single photon peak. The position of the valley significantly changes among devices and pixels since the MaPMT typical gain-uniformity range is $\sim3$. Thus, the threshold must be adjusted channel-by-channel by acquiring the single photon spectra. 
This can be achieved by uniformly illuminating all the pixels and performing a discriminator threshold scan. Figure \ref{fig:SinglePhoton}a shows the results obtained illuminating a MaPMT of the EC with a blue LED and considering the number of signals recorded above each threshold. Taking the derivative of such a curve (fig.\ref{fig:SinglePhoton}b), one obtains the number of events recorded at each threshold level, i.e. the single photon spectrum. In fig.\ref{fig:SinglePhoton}.b the position of both single photon peak and valley is indicated.

\section{Elementary cell - Test beam at CERN}\label{sec:TestBeam}
\begin{figure}[h!]
	\centering
		\begin{minipage}[t]{.4625\textwidth}
			\includegraphics[width=1\textwidth]{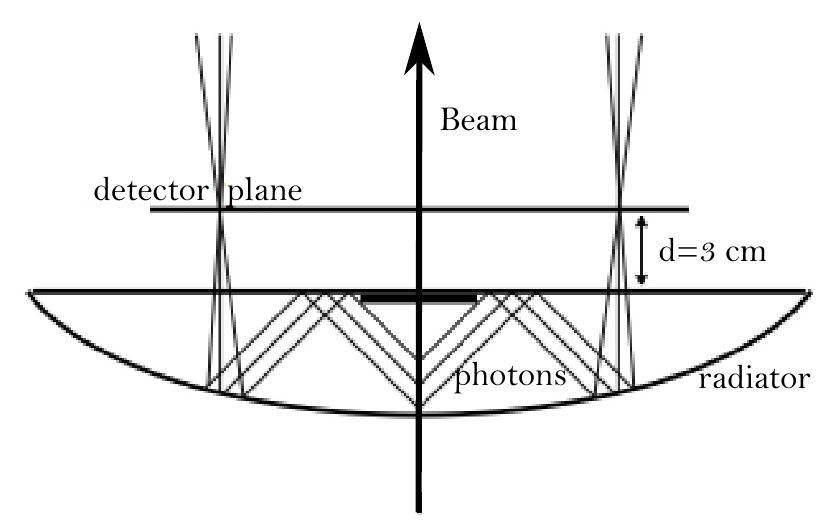}
			\caption{Schematic representation of the plano-convex lens used as Cherenkov radiator in the test-beam. }
			\label{fig:Radiatore}
		\end{minipage}%
	\hspace{10mm}%
		\begin{minipage}[t]{.4625\textwidth}
			\includegraphics[width=1\textwidth]{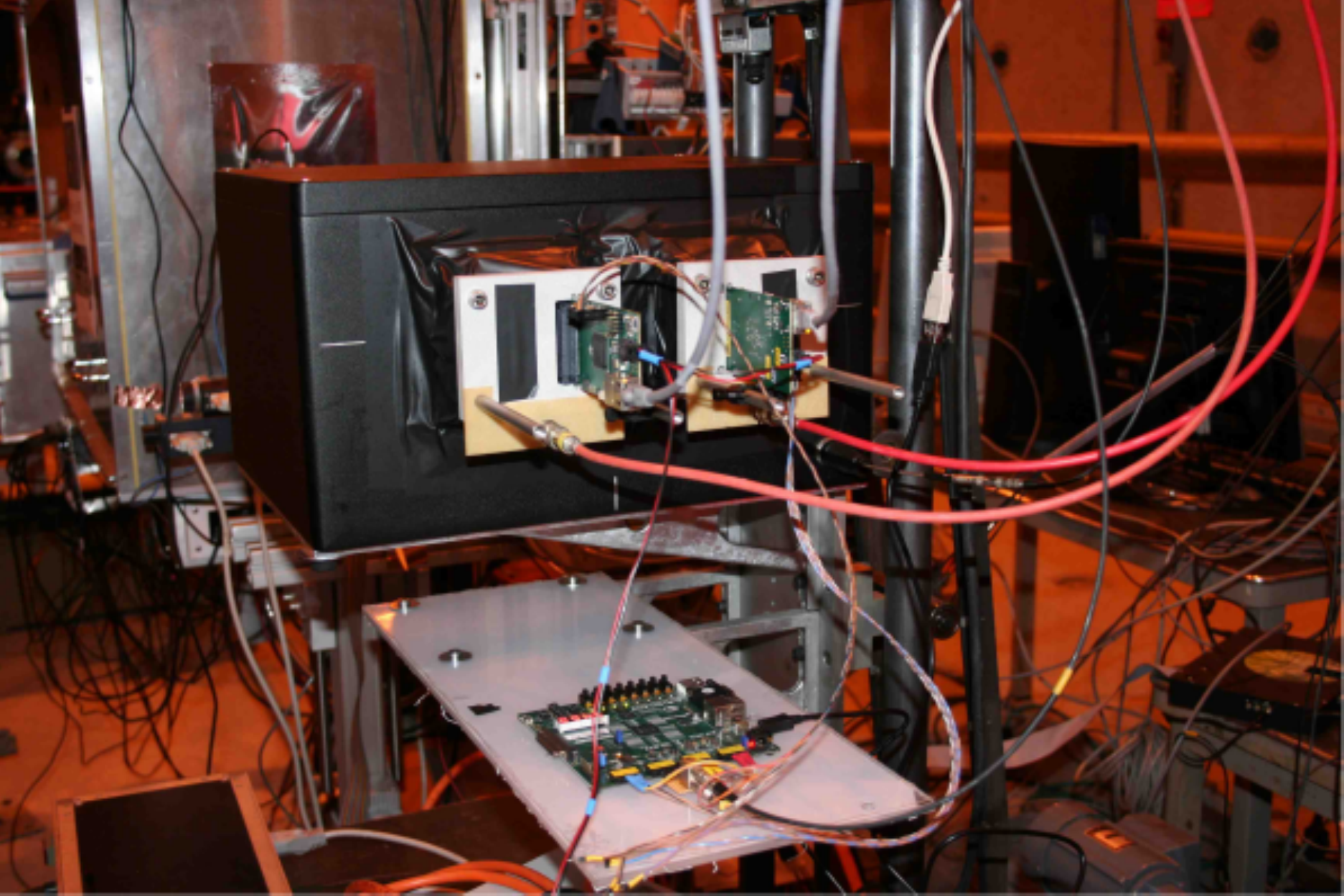}
			\caption{A picture of the setup used for the RICH Upgrade test-beam (October 2014). The light-tight black box and the rear part of the two ECs are visible.}
			\label{fig:testbeamsetup}
		\end{minipage}
\end{figure}

In October and November 2014 the first two complete EC prototypes were studied using the H8 beam line at the SPS North Area. A high energy beam, with well-defined momentum, mainly composed of 180 GeV/c protons and pions was used. The beam provides $\sim10^6\div10^7$ particles per spill. The particles traverse scintillators that provide the trigger to the EC acquisition system. A telescope made of 8 planes of Timepix3 silicon pixel sensors \cite{TimePix3_1} \cite{TimePix3_2} were used to reconstruct the trajectories of the particles. The beam then passes through a light-tight box containing a plano-convex lens, in which Cherenkov photons are emitted. As schematically shown in fig.\ref{fig:Radiatore}, the beam hits the spherical surface of the lens producing Cherenkov photons which are totally reflected from the flat rear surface. The photons are thus focused by the spherical lens to the photodetector plane, about 3 cm distant from the lens output surface. According to the simulation, the expected Cherenkov ring radius is $\sim60$ mm with a spatial resolution of 0.6 mm, limited by the optical system and the MaPMT pixel size. The two EC prototypes are located symmetrically, left and right of the beam center, along the horizontal diameter, so that only two arcs of the Cherekov ring can be detected. A picture of the setup used in the test-beam is shown in fig.\ref{fig:testbeamsetup}.

\begin{figure}[h!]
	\centering
		\begin{minipage}[t]{.4625\textwidth}
			\includegraphics[width=1\textwidth]{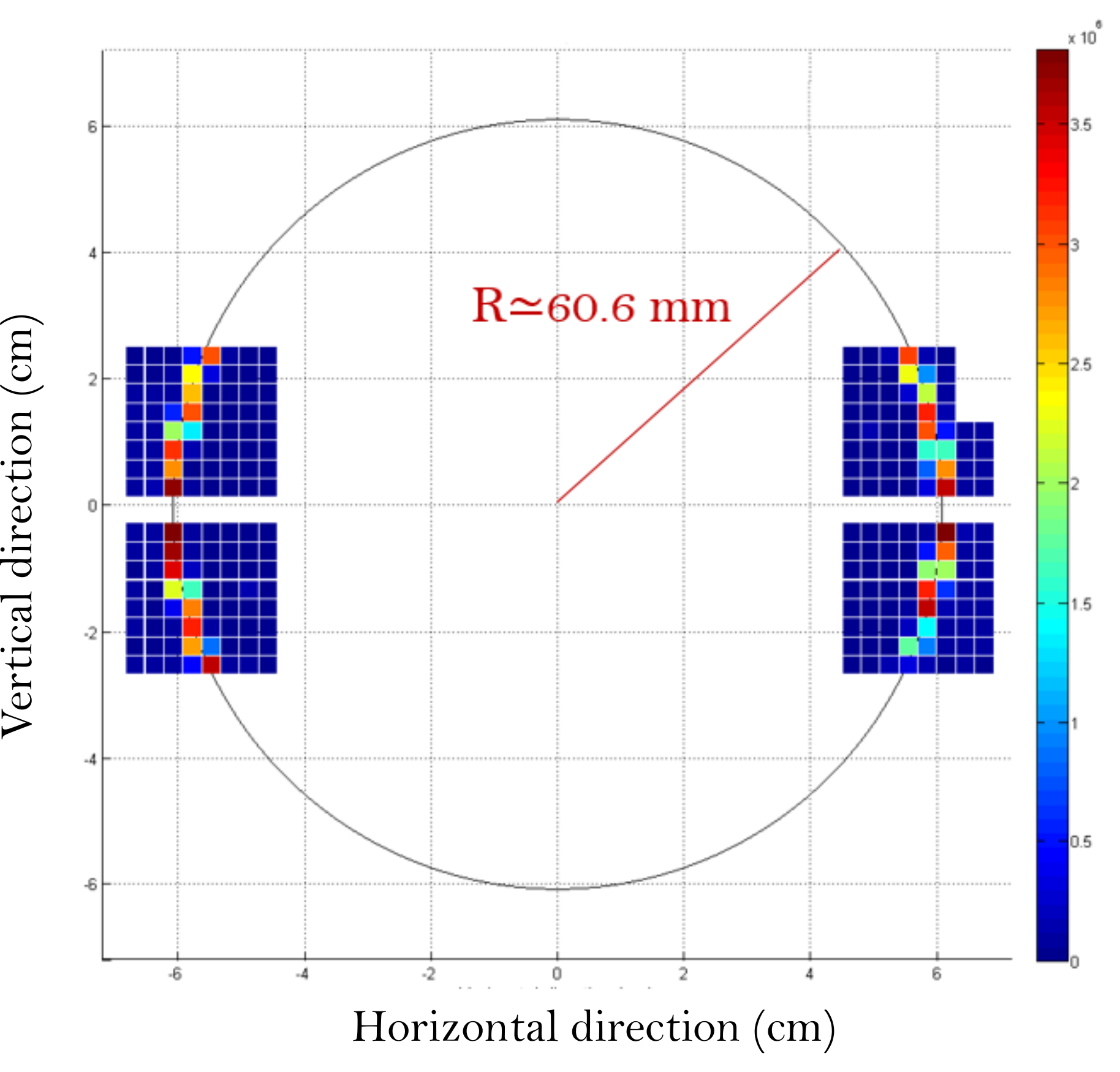}
			\caption{The Cherenkov ring detected in the RICH Upgrade test-beam. The colors of the pixels are related to the occupancy. The fitted Cherenkov ring is also shown.}
			\label{fig:Ring}
		\end{minipage}%
	\hspace{10mm}%
		\begin{minipage}[t]{.4625\textwidth}
			\includegraphics[width=1\textwidth]{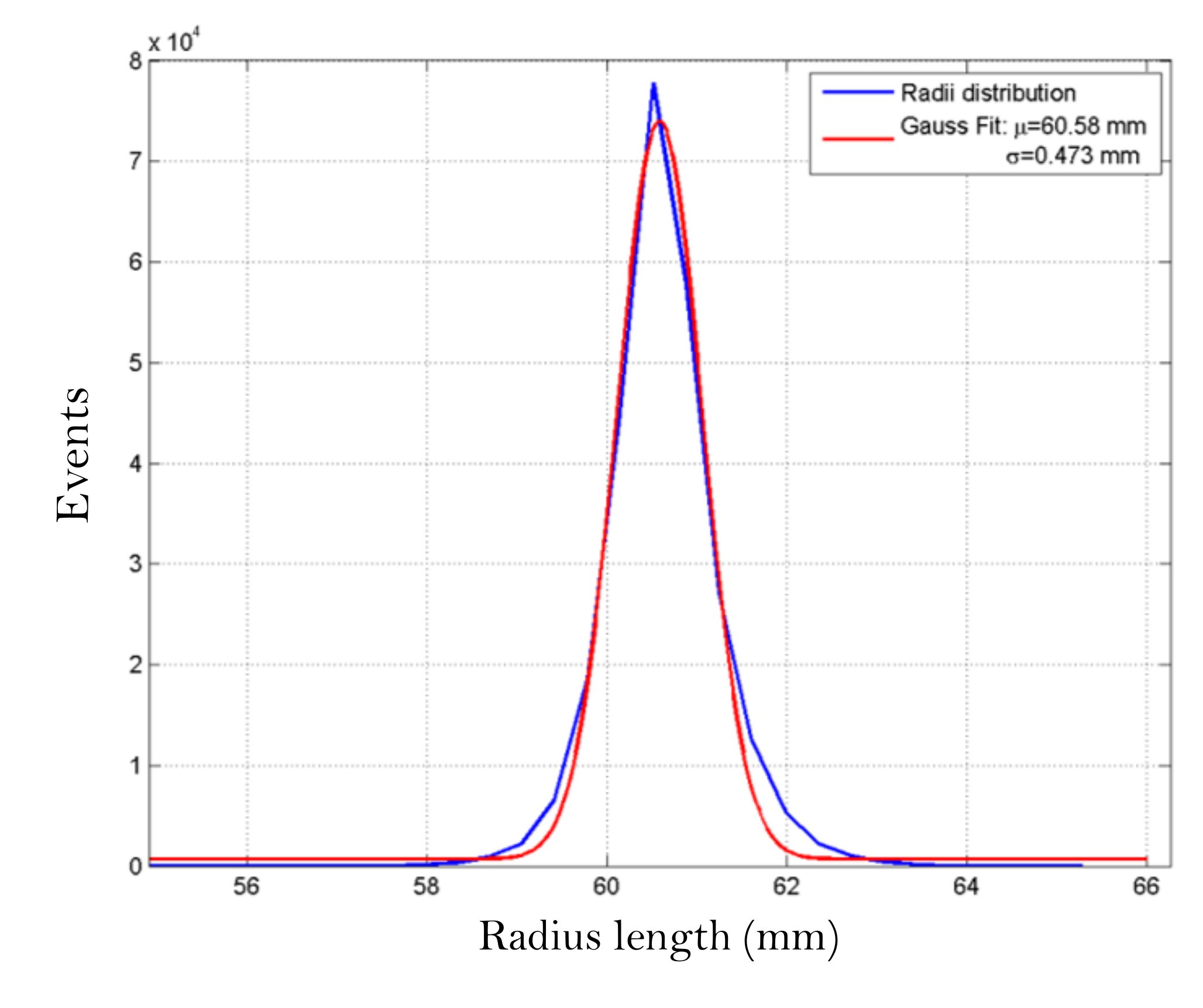}
			\caption{Gaussian shaped Cherenkov ring radius distribution. In agreement with the Monte Carlo simulation, the mean radius value is \hbox{$R\simeq60.6\pm0.5$ mm}.}
			\label{fig:Radii}
		\end{minipage}
\end{figure}

On average four hit pixels per MaPMT were recorded for each event. Figure \ref{fig:Ring} shows the Cherenkov ring obtained by superimposing the signals acquired in $\sim10^5$ events. As it can be seen, the system was able to detect single photons and the radius of the Cherenkov ring was measured to be $\simeq60.6$ mm in agreement with that expected from the simulation. Since we were dealing with a monochromatic beam, the resolution of the Cherenkov radius can be estimated by measuring it for each triggered event and then considering the distribution of radii. Figure \ref{fig:Radii} shows the results: the center of the gaussian distribution represents the average radius value (similar to the one shown in fig.\ref{fig:Ring}) while its width is the spatial resolution. A good spatial resolution of $\sim0.5$ mm was achieved. Similar spatial resolution values can be obtained measuring the position of the beam center along the x-axis, while the position of the ECs with respect to the ring limits the resolution achievable on the y-axis as the ring is almost vertically oriented. 

\newpage
\section{Conclusions}
The Elementary Cell for the LHCb RICH Upgraded detector has been described. It consists of four R11265 MaPMTs read out by a wide-bandwidth, low power consumption, radiation tolerant ASIC, named CLARO. The main features of the chip have been briefly presented. 
Two procedures were developed to calibrate the electronics chain and the photomultipliers in the final RICH environment. Using a DAC it is possible to inject a known charge at the input of the ASIC for its calibration. By performing a discriminator threshold scan, single photon spectra can be obtained for each pixel so that the photosensors can be characterized. The first prototypes were operated in a charged-particle test beam at CERN, where good performance was demonstrated. The Cherenkov ring was detected with a good spatial resolution ($\sim0.5$ mm).


\end{document}